\documentclass[preprint,preprintnumbers,amsmath,amssymb]{revtex4}
\usepackage{graphicx}% Include figure files
\usepackage{bm}% bold math

\begin{document}

\title{Thermal decoherence of long-distance entanglement in spin-$1$ chains}

\author{Xiang Hao}
\altaffiliation{Corresponding author:110523007@suda.edu.cn}

\affiliation{Department of Physics, School of Mathematics and
Physics, Suzhou University of Science and Technology, Suzhou,
Jiangsu 215011, People's Republic of China}

\begin{abstract}

The thermal entanglement is generated by weakly interacting atoms
with an isotropic spin-$1$ chain. The decoherence of the
entanglement is mainly investigated. The effective Hamiltonian is
analytically obtained by the approximation method of perturbation.
The scaling behavior of the effective coupling is numerically
illustrated by the exact diagonalization. It is found out that the
decay of the entanglement is slow in the case of non interacting
spins. The long-distance thermal entangled states can be used as the
noisy channel for the achievement of the quantum teleportation.

PACS: 03.67.Mn, 03.65.Ud, 75.10.Jm, 75.10.Pq
\end{abstract}

\maketitle

\section{Introduction}

The entanglement plays a key role in the main tasks of quantum
information\cite{Nielsen00, Amico07}. In practice, entangled qubits
need be accessed individually for measurements. Consequently, they
are well separated in space. Recently, the long-distance
entanglement \cite{Venuti06, Hartmann06} has been attractive in the
field of quantum information processing. A selected pair of distant
qubits can retain a sizable amount of entanglement at zero
temperature if they are weakly coupled to some spin models. Because
spin chains can serve as an efficient communication channel for
quantum teleportation \cite{Bowen01} and state transfer
\cite{Bose03}, these models are extensively studied. In many schemes
\cite{Ferreira08, Zhu08}, spin-$\frac 12$ Heisenberg chains acted as
the medium for the generation of quantum entanglement when the chain
is kept at the ground state. It is found out that the long-distance
entanglement decreases and vanishes with the length of the gapless
spin chains \cite{Venuti07}. As an appealing spin model, the
spin-$1$ chain exhibits the massive and gapped ground state, which
can be realized through confining an $S=1$ spinor condensate
\cite{Demler02, Yip03} in optical lattices. The quantum
communication in the spin-$1$ chain has been investigated by
\cite{Sanpera07}. Here we expect that the long-distance entanglement
can also be generated by the spin-$1$ chains and show the scaling
property which is different from spin-$\frac 12$ chains. In the
realistic optical lattices, the thermal decoherence from the
temperature is unavoidable \cite{Hofstetter06}. Therefore, it is of
fundamental importance to study the impacts of the thermal noise on
the long-distance entanglement. Using the long-distant entangled
state as the channel, we also suggest the standard scheme of quantum
teleportation.

In this report, the thermal entanglement between a pair of distant
qubits is present when they are weakly coupled to the general
isotropic spin-$1$ chain with bilinear-biquadratic interactions at
finite low temperatures. To study the decoherence, the effective
Hamiltonian between two distant sites is analytically obtained by
the Fr\"{o}hlich transformation \cite{Frohlich1952, Nakajima1953} in
Sec. II. The scaling property of the effective coupling is also
given by the exact diagonalization method. The effects of the
temperature and the relative strength of biquadratic interactions
are considered. In Sec. III, we draw on the master equation to
investigate the decay of the long-distance entanglement. The
protocol of the quantum teleportation is put forward. Finally, a
short discussion concludes the paper.

\section{The effective Hamiltonian at finite low temperatures}

In the optical lattices, a selected pair of two-level atoms $A$ and
$B$ can weakly interact with two open ends of a spin-$1$ chain. At
finite low temperatures, the whole quantum system exhibits the
thermal equilibrium state. To study the time evolution of quantum
states, the total Hamiltonian can be expressed by
\begin{equation}
H=H_{0}+H_{I}=H_{q}+H_{c}+H_{I},
\end{equation}
where
\begin{equation}
H_{q}=\omega(s_A^z+s_B^z),
\end{equation}
describes the intrinsic Hamiltonian of two distant atoms,
\begin{equation}
H_{c}=J\sum_{i=1}^{L-1}[\cos \theta(\vec{S}_{i}\cdot
\vec{S}_{i+1})+\sin \theta (\vec{S}_{i}\cdot \vec{S}_{i+1})^2],
\end{equation}
is the Hamiltonian of general isotropic spin-$1$ chain with the even
length $L$ and
\begin{equation}
H_{I}=J_{p}(\vec{s}_A \cdot \vec{S}_1+\vec{S}_N \cdot \vec{s}_B).
\end{equation}
denotes the weak interaction between two distant atoms and open ends
of the chain. Here $\vec{s}_{A(B)}=\frac 12\sigma_{A(B)}$ and
$\vec{S}_i$ refer to the spin operators of distant atoms and the
$i$th site of the chain respectively. The parameter $\omega$
describes the transition energy from the ground state to the excited
one for each atom, and $J\cos \theta(\sin \theta)$ gives the
strength of the bilinear(biquadratic) coupling. As is well known,
the energy property of the spin model is determined by the angle
$\theta$ \cite{Affleck87}. In the context, the biquadratic coupling
for $|\theta|<\tan^{-1}\frac 13$ need be so weak that the general
ground state of $H_c$ is a total singlet $|\phi_0\rangle$ with the
energy $\epsilon_0$ and the first excited ones are the degenerate
triplet states $|\phi_1^{\lambda=0,1,2}\rangle$ with $\epsilon_1$.
Here the energy gap $\Delta=\epsilon_1-\epsilon_0$ is the famous
Haldane gap. In general, the thermal equilibrium state
$\rho_c(T)=\sum_i\frac{e^{-\epsilon_i/T}}{Z}|\phi_i\rangle\langle
\phi_i|$ where $\epsilon_i$ is the $i$-th eigenvalue of $H_c$ and
$|\phi_i\rangle$ is the corresponding eigenstate. When the low
temperatures satisfy $kT<\Delta$, the components of the ground state
and first excited ones become dominant in the thermal equilibrium
state. For lower temperatures, this assumption of considering just
these states in the thermal fluctuations is more reliable. The
approximate expression of the thermal state can be given by
$\rho_c(T)\simeq \frac {e^{-\epsilon_0/T}}{Z}\left
(|\phi_0\rangle\langle
\phi_0|+e^{-\Delta/T}\sum_{\lambda}|\phi_1^{\lambda}\rangle\langle
\phi_1^{\lambda}|\right )$ where $Z\simeq
e^{-\epsilon_0/T}+3e^{-\epsilon_1/T}$ is the partition function. For
the convenience, the Plank constant $\hbar$ and the Boltzman
constant $k$ are assumed to be one.

In general, the Fr\"{o}hlich transformation \cite{Frohlich1952,
Nakajima1953} is widely used in condensed matter physics. Recently,
this method has been applied to the regime of quantum information
processing \cite{Li05}. As a second-order perturbation
\cite{Frohlich1952, Nakajima1953}, the effective Hamiltonian of the
whole system is $H_{eff}\approx H_{0}+\frac 12[\hat{S},H_{I}]$ where
the anti-Hermitian operator $\hat{S}$ satisfies the relation of
$[H_0,\hat{S}]=H_I$ and the elements of this matrix are given by
$\langle \phi_i^{m}|\hat{S}|\phi_j^{n}\rangle=\frac {\langle
\phi_i^{m}|H_{I}|\phi_j^{n}\rangle}{\epsilon_{i}-\epsilon_{j}},
(i\neq j)$ and the diagonal ones are zero for $m=n,i=j$ \cite{Li05}.
Here $|\phi_{i}^{m}\rangle(m=0,1,\cdots,d_i-1)$ is the energy state
of $H_{c}$ with the corresponding energy $\epsilon_{i}$ and $d_i$ is
the degree of degeneracy. In the case with $J_p\ll J$ at lower
temperature, the spin-$1$ chain is at the state of $\rho_c$ and then
the effective Hamiltonian between two distant atoms is obtained by
\begin{equation}
H_{eff}^{A,B}=\mathrm{Tr_c}\left \{H_{0} \rho_c +\frac
12[\hat{S},H_{I}]\rho_c \right \}.
\end{equation}
where $\mathrm{Tr_c}$ denotes the trace over the complete energy
space of $H_c$. To simplify the calculation, we assume that the
parameter $ \langle
\phi_k^{m}|S^{\alpha}_{i}|\phi_l^{n}\rangle=\tau^{km,ln}_{i,\alpha}=\tau_{i,\alpha}$
where the spin operator $S^{\alpha}=S^{\pm},S^z$. Due to the
invariant symmetry, it is found out that $\sum_{k\neq
l,m,n}\tau_{i,\alpha}\tau^{*}_{j,\beta}/(\epsilon_{k}-\epsilon_{l})=\Omega^{l,\alpha}_{i,j}\delta_{\alpha,\beta}$
for $l=0,1$. Here the sum is always zero if $\alpha \neq \beta$ and
the values $\Omega^{l,\pm}_{i,j}=2\Omega^{l,z}_{i,j}$ are real. As a
consequence, the effective Hamiltonian can be simplified by the
isotropic Heisenberg one
\begin{equation}
H_{eff}^{A,B}=J_{eff}\vec{s}_A \cdot \vec{s}_B+H_q+C.
\end{equation}
Here the constant of $C$ is irrelevant to the long distant
entanglement. The effective Heisenberg coupling $J_{eff}=-\frac
{2J_{p}^2e^{-\epsilon_0/T}}{Z}(\Omega_{1,L}^{0,z}+e^{-\Delta/T}\Omega_{1,L}^{1,z})$
is closely dependent on the energy property of $H_0$. By means of
the exact diagonalization method, the scaling property of $J_{eff}$
at finite low temperatures is demonstrated in Fig. 1. It is shown
that the values increase almost exponentially and arrive at a steady
one with the length of the chain. According to \cite{Venuti07}, the
effective coupling $J_{eff}$ is mainly determined by the
singlet-triplet gap of the whole system $H$. From the numerical
results of \cite{White93}, the gap of $H$ for $L\sim20$ is almost
the steady one. Therefore, the values of $J_{eff}$ saturate rapidly
with the increase of the length. This means that the effective
coupling can be obtained at finite low temperatures when distant
sites are taken infinitely far away. Notice that the parameter
$\Omega^{l,\alpha}_{i,j}$ must be calculated by all of eigenvectors
of $H_c$ and $\rho_c$ is approximately expressed in the
singlet-triplet subspace. For a simplest example of $L=2$, the
Hamiltonian of $H_c$ can be expanded by
\begin{equation}
H_c=\sum_{\lambda=0}^{2}\epsilon_{\lambda}\hat{P}_{\lambda}
\end{equation}
where the projectors
$\hat{P}_{\lambda}=\sum_{S^z_{tot}}|S_{tot}=\lambda,S^z_{tot}\rangle
\langle S_{tot}=\lambda,S^z_{tot}|$ and
$S^z_{tot}=-\lambda,\cdots,\lambda$. For very small
$|\theta|<\tan^{-1}\frac 13$, the energy spectrum is given by the
ground energy $\epsilon_0=-2J(\cos\theta-2\sin \theta)$, the first
excited one $\epsilon_1=-J(\cos\theta-\sin \theta)$ and the second
$\epsilon_2=J(\cos\theta+\sin \theta)$. Thus the effective coupling
is analytically written by
\begin{equation}
J_{eff}=\frac {e^{-\epsilon_0/T}}{3Z}\left (\frac
{4J_p^2-4J_p^2e^{-\Delta/T}}{\epsilon_1-\epsilon_0}-\frac
{5J_p^2e^{-\Delta/T}}{\epsilon_2-\epsilon_1} \right ).
\end{equation}
It is necessary to consider the effects of the temperatures and
relative strength of biquadratic coupling $\theta$ on the effective
coupling. From Fig. 2, it is seen that the values $J_{eff}$ are
increased by the slight increase of $\theta$. For the even length of
the chain, the parameters $\Omega^{l,\alpha}_{1,L}(l=0,1)$ are
negative. In accordance with Eq.(8), the values of $J_{eff}$ can be
enhanced slightly because the bigger angle $\theta$ leads to the
smaller energy gap $\Delta$. For the low temperature, the effective
coupling is mainly determined by the first item of $\frac
{2J_{p}^2e^{-\epsilon_0/T}}{Z}|\Omega_{1,L}^{0,z}|=\frac
{|\Omega_{1,L}^{0,z}|}{1+3e^{-\Delta/T}}$ which is decreased with
increasing the temperature.

\section{Decoherence of entanglement in thermal noise}

The state of two distant atoms $\rho^{A,B}$ can be gained by tracing
out the variables of the chain from the thermal state of the whole
system. However, if the temperature $kT\ll \Delta$, we do not expect
real excitations of the spin chain to be present \cite{Ferreira08}.
Only the subspace of the states described by $H^{A,B}_{eff}$ will be
populated and then we can calculate the correlations between two
atoms using $\rho^{A,B}=e^{-H^{A,B}_{eff}/T}/Z_q$ where
$Z_q=\mathrm{Tr}[e^{-H^{A,B}_{eff}/T}]$. When two distant sites are
simultaneously coupled to the chain, the thermal state $\rho^{A,B}$
can be generated. In accordance with
\cite{Wang02,Bayat05,Nielsen,Arnesen01,Wang01}, the concurrence of
$\rho^{A,B}$ can be written by $C=\frac 1{Z_q} \max \{0,
e^{3J_{eff}/4T}-3e^{-J_{eff}/4T} \}$. Therefore, thermal
entanglement exists if the effective coupling satisfies $\frac
{J_{eff}}{T}>\ln3$. From the point of view of practice, the local
operations concerning two distant entangled atoms are needed. It is
reasonable to assume that two atoms are coupled with its local
thermal reservoirs $E_A,E_B$. According to \cite{Yu04}, the two
independent reservoirs can lead to the local decoherence of
entanglement. Suppose that the initial state at $t=0$ is
$\rho_{tot}=\rho^{A,B}\otimes
(|0_{E_A}0_{E_B}\rangle\langle0_{E_A}0_{E_B}|)$ where
$|0_{E_A}0_{E_B}\rangle$ denotes the vacuum state of the two local
reservoirs. The evolution of quantum state between atoms $A$ and $B$
is given by the master equation
\begin{equation}
\dot{\rho}(t)=-i[H_{eff},\rho]+\hat{L}(\rho),
\end{equation}
where the Lindbald operator
\begin{align}
\hat{L}(\rho)=\sum_{i=A,B}
&(\bar{n}_i+1)\Gamma_i(2\sigma_i^{-}\rho\sigma_i^{+}-\rho\sigma_i^{+}\sigma_i^{-}-\sigma_i^{+}\sigma_i^{-}\rho)&
\nonumber \\
&+\bar{n}_i\Gamma_i(2\sigma_i^{+}\rho\sigma_i^{-}-\rho\sigma_i^{-}\sigma_i^{+}-\sigma_i^{-}\sigma_i^{+}\rho).&
\end{align}
Here $\bar{n}_i=\bar{n}$ is the mean number of the thermal reservoir
and $\Gamma_i=\Gamma$ signifies the rate of spontaneous emission for
each atom.

If one of two weak couplings $J_p$ is turned off after the
preparation of the long-distance entanglement, the effective
Hamiltonian of two atoms is obtained by $H_{eff}=H_q+C^{'}_{eff}$
which means there is no mutual interaction between atoms. In this
case, the evolution of $\rho(t)$ can be described by a completely
positive trace-preserving map \cite{Aolita08}. For a general
two-qubit mixed state
$\rho(0)=\sum_{kl,mn}a_{mn,kl}|kl\rangle_{AB}\langle mn|$, the
evolved state in time can be written by
$\rho(t)=\sum_{kl,mn}\sum_{j,j'}a_{mn,kl}(K_{Aj}|k\rangle_A\langle
m|K^{\dag}_{Aj})\otimes(K_{Bj'}|l\rangle_B\langle l|K^{\dag}_{Bj'})
$ where the Kraus operators $K_{i0}=\sqrt{\frac
{\bar{n}+1}{2\bar{n}+1}}(|g\rangle_i\langle
g|+\sqrt{1-p}|e\rangle_i\langle e|)$, $K_{i1}=\sqrt{\frac
{(\bar{n}+1)p}{2\bar{n}+1}}|g\rangle_i\langle e|$,
$K_{i2}=\sqrt{\frac
{\bar{n}}{2\bar{n}+1}}(\sqrt{1-p}|g\rangle_i\langle
g|+|e\rangle_i\langle e|)$ and $K_{i3}=\sqrt{\frac
{\bar{n}p}{2\bar{n}+1}}|e\rangle_i\langle g|$. Here $|g(e)\rangle_i$
is the ground(excited) state of atoms $i=A,B$ and $p(t)=1-e^{\frac
{\Gamma(2\bar{n}+1)t}{2}}$ means the probability of the atom
exchanging a quantum with the reservoir. The density matrix of the
quantum state at any time is expanded in the Hilbert space of
$\{|gg\rangle_{AB},|ge\rangle_{AB},|eg\rangle_{AB},|ee\rangle_{AB}\}$
\begin{equation}
\rho(t)=\frac {1}{Z_q}\left(\begin{array}{cccc}
            u&0&0&0\\
            0&x&y&0\\
            0&y&x&0\\
            0&0&0&v
            \end{array}\right).
\end{equation}
The elements of $\rho(t)$ are expressed by $u=(1-a)^2e^{-(\frac
{J_{eff}}{4}-\omega)/T}+a^2e^{-(\frac
{J_{eff}}{4}+\omega)/T}+a(1-a)(e^{-\frac {J_{eff}}{4T}}+e^{\frac
{3J_{eff}}{4T}})$, $v=(1-a)^2e^{-(\frac
{J_{eff}}{4}+\omega)/T}+a^2e^{-(\frac
{J_{eff}}{4}-\omega)/T}+a(1-a)(e^{-\frac {J_{eff}}{4T}}+e^{\frac
{3J_{eff}}{4T}})$, $x=a(1-a)(e^{-(\frac
{J_{eff}}{4}+\omega)/T}+e^{-(\frac {J_{eff}}{4}-\omega)/T})+\frac
12[(1-a)^2+a^2](e^{-\frac {J_{eff}}{4T}}+e^{\frac {3J_{eff}}{4T}})$
and $y=\frac {1-p}{2}(e^{-\frac {J_{eff}}{4T}}-e^{\frac
{3J_{eff}}{4T}})$ where $a=\frac {\bar{n}p}{2\bar{n}+1}$. The
concurrence \cite{Wang02,Bayat05,Nielsen,Arnesen01,Wang01} is used
to evaluate the long distant entanglement
\begin{equation}
C=\frac 2{Z_q}\max \{0, |y|-\sqrt{uv} \}.
\end{equation}
On the other hand, it is assumed that the two atoms directly
interact with each other in the form of the Hamiltonian given by
Eq.(6). In this case, the analytical solution of the master equation
is tedious. The expression of the density matrix of quantum states
is also similar to that of Eq.(11). The decoherence of the thermal
entanglement in two cases can be illustrated by Fig. 3(a). It is
seen that the entanglement of two qubits without mutual interactions
is decreased much more slowly than that of two directly interacting
qubits. This point demonstrates that the decoherence time for long
distant entanglement is so long as to be useful for the
implementation of solid-state quantum computation.

The standard teleportation through the mixed states can be regarded
as a general depolarising channel \cite{Bowen01}. An arbitrary
unknown quantum state $|\Psi\rangle=\cos \frac
{\theta}2|g\rangle+\sin \frac
{\theta}2e^{i\varphi}|e\rangle,(0\leq\theta\leq \pi,0\leq\varphi\leq
2\pi)$ is destroyed and its replica state appears at the remote
place after applying the Bell measurement and the corresponding
local operations. When single-qubit state $\rho_{in}=|\Phi\rangle
\langle \Psi|$ is teleported via the noisy channel of $\rho$ like
Eq.(12), the output state $\rho_{out}$ is written by
\begin{equation}
\rho_{out}=\sum_i \mathrm{Tr}[E^i\rho]\sigma^i\rho_{in}\sigma^i.
\end{equation}
In the above equation, $i=0,x,y,z$ and the projectors
$E^{0}=|\psi^{-}\rangle\langle\psi^{-}|,E^{i}=\sigma^iE^0\sigma^i$
where a Bell state $|\psi^{-}\rangle=\frac 1{\sqrt
2}(|ge\rangle-|eg\rangle)$. The average fidelity of this
teleportation is given by
\begin{equation}
F_{A}=\frac {\displaystyle \int_{0}^{2\pi}d\phi\!\int_{0}^{\pi}
F\,\sin\theta\,d\theta} {4\pi}=\frac 16+\frac {3x-2y}{3Z_q}
\end{equation}
According to \cite{Nielsen00}, the fidelity for a pure input state
$F=\{\mathrm{Tr}[\sqrt{(\rho_{in})^{1/2}\rho_{out}(\rho_{in})^{1/2}}]\}^{2}=\mathrm{Tr}[\rho_{out}\rho_{in}]$.
The effect of the thermal noise on the average fidelity of the
standard teleportation is illustrated by Fig. 3(b). It is shown that
the average fidelity of quantum teleportation with thermal
decoherence is larger than $2/3$ before a certain time. This means
that the quantum teleportation via the channel of long-distance
entangled state is better than the classic communication in the
range of finite time. In the condition of the thermal noise, the
quantum teleportation as the channel of the long-distant thermal
entangled state is better than that of the thermal entangled state
between two qubits interacting directly.

\section{Discussion}

The long-distance thermal entanglement can be obtained when two
atoms are weakly interacting with the isotropic spin-$1$ chain at
finite low temperatures. For the massively gapped quantum systems,
the scaling law for the effective coupling shows the exponential
increase with the length of the spin chain. Under the influence of
thermal noise, the entanglement of two distant qubits without mutual
interactions is decreased much more slowly. It is demonstrated that
the resource of long-distance entanglement can be used for quantum
information processing. We suggest the efficient scheme of the
standard teleportation via the channel of long-distance
entanglement.

\section{Acknowledgement}

X.H. was supported by the Initial Project of Research in SUST and
the National Natural Science Foundation of China No. 10774108.

\newpage

{\Large Figure caption}

Figure 1

The finite-size scaling behavior of the effective coupling
$J_{eff}/J$ is demonstrated at the low temperature $T=0.1$. The
square data were obtained with the exact diagonalization method when
the couplings $J_p/J=0.1$ and $\theta=0$. The solid line was the
exponential fit curve for the data.

Figure 2

The effective coupling $J_{eff}/J$ is plotted as functions of the
temperature $T$ and the relative strength of the biquadratic
coupling $\theta$ when the couplings $J_p/J=0.1$.

Figure 3

(a). The decoherence of the entanglement is shown. (b). The average
fidelity of the standard teleportation is illustrated. The dash line
denotes the case that two qubits interact directly. The solid one
refers to the case that one of the weak couplings between the
distant atoms and the chain is turned off. The length of the chain
is infinite and $J_p/J=0.1$, $\theta=0$ and $T=0.01$. The mean
number of the thermal reservoir is chosen to be $\bar{n}=1$ and the
rate of spontaneous emission is $\gamma=0.1$.


\begin{references}
\bibitem{Nielsen00} M. A. Nielsen and I. L. Chuang, {\it Quantum
Computation and Quantum Information}(Cambridge University Press,
Cambridge, 2000).
\bibitem{Amico07} L. Amico, R. Fazio, A. Osterloh, and V. Vedral,
e-print arXiv:quant-ph/070344v1.
\bibitem{Venuti06} L. Campos Venuti, C. Degli Esposti Boschi, and M.
Roncaglia, Phys. Rev. Lett. \textbf{96}, 247206(2006).
\bibitem{Hartmann06} J. Hartmann, M. E. Reuter, and M. B. Plenio,
New J. Phys. \textbf{8}, 94(2006).
\bibitem{Bowen01} G. Bowen and S. Bose, Phys. Rev. Lett.
\textbf{87}, 267901(2001).
\bibitem{Bose03} S. Bose, Phys. Rev. Lett. \textbf{91},
207901(2003).
\bibitem{Ferreira08} A. Ferreira and J. M. B. Lopes dos Santos,
Phys. Rev. A\textbf{77}, 034301(2008).
\bibitem{Zhu08} X. Hao and S. Zhu, Phys. Rev. A\textbf{78},
044302(2008).
\bibitem{Venuti07} L. Campos Venuti, C. Degli Esposti Boschi, and M.
Roncaglia, Phys. Rev. Lett. \textbf{99}, 060401(2007).
\bibitem{Demler02} E. Demler and F. Zhou, Phys. Rev. Lett.
\textbf{88}, 163001(2002).
\bibitem{Yip03} S. K. Yip, Phys. Rev. Lett. \textbf{90},
250402(2003).
\bibitem{Sanpera07} O. Romero-Isart, K. Eckert, and A. Sanpera,
Phys. Rev. A\textbf{75}, 050303(R)(2007).
\bibitem{Hofstetter06}  W. Hofstetter, Philos. Mag. \textbf{86},
1891(2006).
\bibitem{Frohlich1952} H. Fr\"{o}hlich, Proc. Roy. Soc. A \textbf{215},
291(1952).
\bibitem{Nakajima1953} S. Nakajima, Adv. Phys. \textbf{3},
325(1953).
\bibitem{Affleck87} I. Affleck, T. Kennedy, E. H. Lieb, and H.
Tasaki, Phys. Rev. Lett. \textbf{59}, 799(1987).
\bibitem{Li05} Y. Li, T. Shi, B. Chen, Z. Song and C. P. Sun, Phys.
Rev. A\textbf{71}, 022301(2005).
\bibitem{White93} S. R. White and D. A. Huse, Phys. Rev.
B\textbf{48}, 3844(1993).
\bibitem{Wang02} X. Wang and P. Zanardi, Phys. Lett. A\textbf{301},
1(2002).
\bibitem{Bayat05} A. Bayat and V. Karimipour, Phys. Rev. A\textbf{71},
042330(2005).
\bibitem{Nielsen} M. A. Nielsen, Ph.D Thesis, quant-ph/0011036.
\bibitem{Arnesen01} M. C. Arnesen, S. Bose and V. Vedral, Phys. Rev. Lett. \textbf{87}, 017901(2001).
\bibitem{Wang01} X. Wang, et. al., J. Phys. A:Math. Gen. \textbf{34},
11307(2001).
\bibitem{Yu04} T. Yu and J. H. Eberly, Phys. Rev. Lett. \textbf{93},
140404(2004).
\bibitem{Aolita08} L. Aolita, R. Chaves, D. Cavalcanti, A. Ac\'{i}n,
and L. Davidovich, Phys. Rev. Lett. \textbf{100}, 080501(2008).


\end{references}
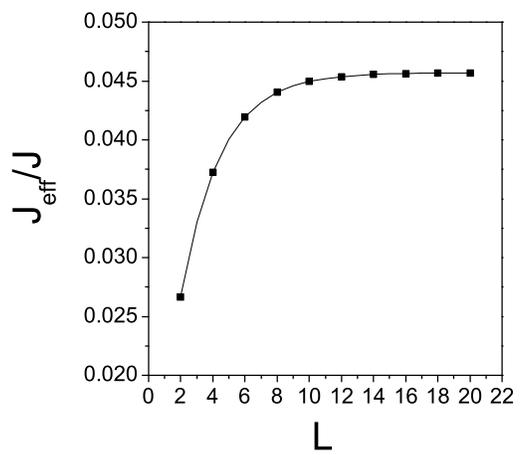
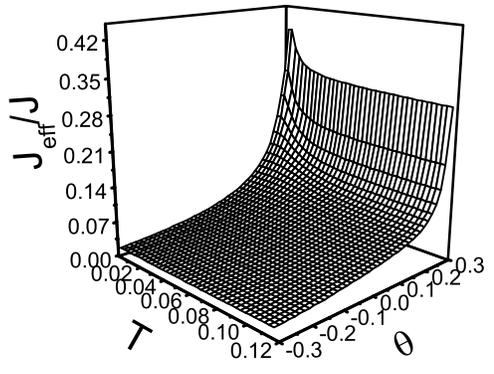
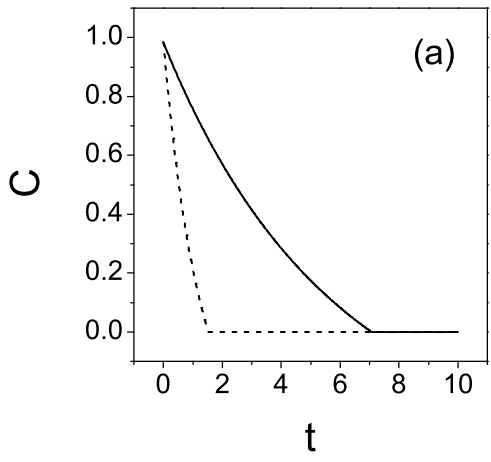
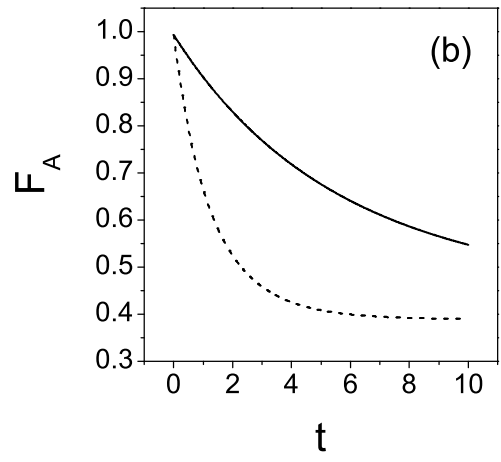
\end{document}